\documentclass[twocolumn,showpacs,aps,prl,superscriptaddress]{revtex4}

\usepackage{graphicx}
\usepackage{dcolumn}
\usepackage{amsmath}
\usepackage{units}

\input{pubboard/babarsym}

\newcommand{\etal}{\emph{et. al.}}
\newcommand{\microns}{\ensuremath{\mu{\rm m}}}
\newcommand{\Btag}{\ensuremath{\B_\mathrm{tag}}}
\newcommand{\Bztojpsiks}{\ensuremath{\Bz\to\jpsi\KS}}
\newcommand{\Bztophiks} {\ensuremath{\Bz\to\phi\KS}}
\newcommand{\Bztoetapks} {\ensuremath{\Bz\to\eta'\KS}}
\newcommand{\sintwobeta} {\ensuremath{\sin 2\beta}}
\newcommand{\Bztokpkmks} {\ensuremath{\Bz\to K^+ K^- \KS}}
\newcommand{\Bztokspiz} {\ensuremath{\Bz \to \KS\piz}}
\newcommand{\ckspiz} {\ensuremath{C_{\KS\piz}}}
\newcommand{\skspiz} {\ensuremath{S_{\KS\piz}}}
\def\cf {\ensuremath{C_f}}
\def\sf {\ensuremath{S_f}}
\newcommand{\fish}    {\ensuremath{\cal F}}

\newcommand{\zrec}{\ensuremath{z_{\CP}}}
\newcommand{\ztag}{\ensuremath{z_\mathrm{tag}}}

\newcommand{\BABARPubYear}    {04}
\newcommand{\BABARPubNumber}  {005}
\newcommand{\SLACPubNumber} {10363}
\newcommand{\LANLNumber} {xxxxxxx}

\def\figurebox#1#2#3{%
    \def\arg{#3}%
    \ifx\arg\empty
    {\hfill\vbox{\hsize#2\hrule\hbox to #2{\vrule\hfill\vbox to #1{\hsize#2\vfill}\vrule}\hrule}\hfill}%
    \else
    {\hfill\epsfbox{#3}\hfill}%
    \fi}

\long\def\inst#1{\par\nobreak\kern 4pt\nobreak
    {\it #1}\par\vskip 10pt plus 3pt minus 3pt}

\begin{document}

\preprint{\babar-PUB-\BABARPubYear/\BABARPubNumber}
\preprint{SLAC-PUB-\SLACPubNumber}

\begin{flushleft}
  \babar-PUB-\BABARPubYear/\BABARPubNumber\\
  SLAC-PUB-\SLACPubNumber\\
  hep-ex/\LANLNumber\\[10mm]
\end{flushleft}

\title{
  { \Large \bf \boldmath Measurements of \CP-violating Asymmetries in
    \Bztokspiz\ Decays }
}

%
\author{B.~Aubert}
\author{R.~Barate}
\author{D.~Boutigny}
\author{F.~Couderc}
\author{J.-M.~Gaillard}
\author{A.~Hicheur}
\author{Y.~Karyotakis}
\author{J.~P.~Lees}
\author{V.~Tisserand}
\author{A.~Zghiche}
\affiliation{Laboratoire de Physique des Particules, F-74941 Annecy-le-Vieux, France }
% ...
\collaboration{The \babar\ Collaboration}
\noaffiliation


\date{\today}

\begin{abstract}
  We present a measurement of the time-dependent \CP-violating (CPV)
  asymmetries in \Bztokspiz\ decays based on 124 million $\Y4S\to\BB$
  decays collected with the \babar\ detector at the PEP-II
  asymmetric-energy $B$ Factory at SLAC. In a sample containing
  $122\pm 16$ signal decays, we obtain the magnitude of the direct CPV
  asymmetry $\ckspiz = 0.40^{+0.27}_{-0.28} \pm 0.09$ and the
  magnitude of the CPV asymmetry in the interference between mixing
  and decay $\skspiz = 0.48^{+0.38}_{-0.47} \pm 0.06$ where the first
  error is statistical and the second systematic.
\end{abstract}


\maketitle

The \babar~\cite{BaBarSin2betaObs} and Belle~\cite{BelleSin2betaObs}
collaborations recently reported observation of \CP\ violation in $B$
meson decays through measurements of the time-dependent \CP-violating
(CPV) asymmetry in \Bz\ decays into charmonium final states. In the
framework of the Standard Model (SM), where \CP\ violation is a
consequence of the presence of a complex phase in the
Cabibbo-Kobayashi-Maskawa (CKM) quark mixing matrix~\cite{CKM}, these
measurements determine the parameter \sintwobeta, with
$\beta\equiv\arg (-V_{cd}V^*_{cb}/V_{td}V^*_{tb})$. The consistency of
the observed value of \sintwobeta\ with the Standard Model
expectations provides strong evidence that the CKM mechanism is the
dominant source of \CP\ violation in the quark sector. A major goal of
the experimental studies of $B$ decays is to provide additional
information to examine the validity of this conclusion and search for
evidence of new physics (NP) in possible deviations from the SM.  One
avenue for the observation of NP is provided by \CP{} violation
studies of decays dominated by penguin loop-level $b \to s\qbar\q$
$(\q=\{d,s\})$
transitions~\cite{Grossman:1996ke,Ciuchini:1997zp,ref:cc}.  While in
the SM the time-dependent CPV asymmetries in these decays measure
\sintwobeta{}, additional radiative loop contributions from NP
processes may alter this expectation.  Presently, the \B{} factory
experiments have explored time-dependent CPV asymmetries in three such
decays, which in the SM are dominated by the penguin $\b\to\s\sbar\s$
transition: \Bztoetapks{}~\cite{Abe:2003yt,Aubert:2003bq},
\Bztokpkmks{}~\cite{Abe:2003yt}, and
\Bztophiks{}~\cite{Abe:2003yt,ref:browderlp03}.  The latter results
hint at a possible deviation from the SM, but are inconclusive.

In this letter we present the first measurement of the time-dependent
CPV asymmetries in the decay \Bztokspiz, which has a measured
branching fraction $\BR(\Bztokspiz)= (11.9\pm 1.5) \cdot
10^{-6}$~\cite{ref:HFAG}.  The CKM and color suppression of the
tree-level $b\to s\bar{u}u$ transition leads to the expectation that
this decay is dominated by a top quark mediated $\b\to\s\dbar\d$
penguin diagram, which carries a weak phase
$\arg(V_{\t\b}V_{\t\s}^*)$. If other contributions, such as the
$\b\to\s\u\ubar$ tree amplitude, are ignored, the time-dependent CPV
asymmetry is governed by \sintwobeta. The deviation from \sintwobeta{}
due to standard model contributions with a different weak phase is
estimated to be at most $0.2$~\cite{Gronau:2003kx}.

The results presented here are based on $124$ million $\Y4S\to\BB$
decays collected in 1999-2003 with the \babar\ detector at the PEP-II
$\epem$ collider, located at the Stanford Linear Accelerator Center.
The \babar\ detector, which is fully described in~\cite{ref:babar},
provides charged particle tracking through a combination of a
five-layer double-sided silicon micro-strip detector (SVT) and a
40-layer central drift chamber (DCH), both operating in a
\unit[1.5]{T} magnetic field in order to provide momentum
measurements. Charged kaon and pion identification is achieved through
measurements of particle energy-loss ($dE/dx$) in the tracking system
and Cherenkov cone angle ($\theta_c$) in a detector of internally
reflected Cherenkov light (DIRC).  A segmented CsI(Tl) electromagnetic
calorimeter (EMC) provides photon detection and electron
identification.  Finally, the instrumented flux return (IFR) of the
magnet allows discrimination of muons from pions.

We search for \Bztokspiz\ decays in hadronic events, which are
selected based on charged particle multiplicity and event
topology~\cite{ref:Sin2betaPRD}.  We reconstruct $\KS\to\pip\pim$
candidates from pairs of oppositely charged tracks. The two-track
combinations must form a vertex with $\pip\pim$ invariant mass within
$3.5\sigma$ of the nominal \KS\ mass~\cite{Hagiwara:fs} and
reconstructed proper lifetime greater than five times its uncertainty.
We form $\piz\to\gamma\gamma$ candidates from pairs of photon
candidates in the EMC that are isolated from any charged tracks, carry
a minimum energy of \unit[30]{\mev}, and possess the expected lateral
shower shapes.  Finally, we construct \Bztokspiz{} candidates by
combining \KS{} and \piz{} candidates in the event.  For each $B$
candidate two nearly independent kinematic variables are computed,
namely the energy-substituted mass $\mes=\sqrt{(s/2+{\bf p}_i{\bf
    p}_B)^2/E_i^2+p^2_B}$, and the energy difference
$\DeltaE=E^*_B-\sqrt{s}/2$. Here, $(E_i,{\bf p}_i)$ is the
four-vector of the initial \epem{} system, $\sqrt{s} = \sqrt{
  E_i^2-p_i^2}$ is the center-of-mass energy, ${\bf p}_B$ is the
reconstructed momentum of the \Bz{} candidate and $E_B^*$ is its
energy calculated in the \epem{} rest frame. For signal decays, the
\mes{} distribution peaks near the \Bz{} mass with a resolution of
\unit[$\sim3.1$]{\mevcc} and the \DeltaE{} distribution peaks near
zero with a resolution of \unit{$\sim 40$}{\mev}. Both the \mes{} and
the \DeltaE{} distribution exhibit a low-side tail from energy leakage
out of the EMC.  We select candidates within the window
\unit[$5.2<\mes<5.29$]{\gevcc} and \unit[$-150<\DeltaE<150$]{\mev},
which includes the signal peak and a ``sideband'' region for
background characterization. For the~\unit[$1.7$]{\%} of events
with more than one candidate we select the combination with the
smallest $\chi^2=\sum_{i=\piz,\KS} (m_i-m'_i)^2/\sigma^2_{m_i}$, where
$m_i$ ($m'_i$) is the measured (nominal) mass and $\sigma_{m_i}$ is
the estimated uncertainty on the mass of particle $i$.

For each \Bztokspiz{} candidate we examine the remaining tracks and
neutral candidates in the event to determine if the other \B{} meson,
\Btag{}, decayed as a \Bz{} or a \Bzb{} (flavor tag).  Time-dependent
CPV asymmetries are determined by reconstructing the distribution of
the difference of the proper decay times, $\deltat\equiv
t_{\CP}-t_\text{tag}$, where the $t_{\CP}$ refers to the signal \Bz{}
and $t_\text{tag}$ to the other \B{}. At the $\Upsilon(4S)$ resonance,
the $\deltat$ distribution follows
\begin{eqnarray}
  \label{eqn:td} 
  \lefteqn{{\cal P}^{\Bz}_{\Bzb}(\deltat) \; = \; \frac{e^{-|\deltat|/\tau}}{4\tau} \times }\; \\
   && \left[ \: 1 \; \pm \; 
    \left( \: S_f \sin{( \deltamd\deltat)} - C_f \cos{( \deltamd\deltat)} \: \right) \: \right] \; , \nonumber
\end{eqnarray}
where the upper (lower) sign corresponds to \Btag{} decaying as \Bz{}
(\Bzb), $\tau$ is the \Bz{} lifetime averaged over the two mass
eigenstates, \deltamd{} is the mixing frequency, $C_f$ is the
magnitude of direct CPV in the decay to final state $f$ and $S$ the
magnitude of CPV in the interference between mixing and decay. For the
case of pure penguin dominance, we expect $S_{\KS\piz}=\sin2\beta$,
and $C_{\KS\piz}=0$.

We extract the CPV parameters from an unbinned maximum-likelihood fit
to kinematic, event shape, flavor tag, and time structure variables.
We verified that the selected observables are sufficiently independent
that we can construct the likelihood from the product of one
dimensional probability density functions (PDFs).  The PDFs for signal
events are parameterized from either more copious fully-reconstructed
$B$ decays in data or from simulated samples.  For background PDFs we
select the functional form from data in the sideband regions of the
other observables where backgrounds dominate.  We include these
regions in the fitted sample and simultaneously extract the parameters
of the background PDFs along with the CPV measurements.

The sample of \Bztokspiz{} candidates is dominated by random $\KS\piz$
combinations from $\epem\to\qqbar$ $(\q=\{u,d,s,c\})$ fragmentation.
Monte Carlo studies show that contributions from other \B{} meson
decays can be neglected. We exploit topological observables to
discriminate the jet-like $\epem\to\qqbar$ events from the more
uniformly distributed \BB{} events. In the $\Upsilon(4S)$ rest frame
we compute the angle $\theta^*_S$ between the sphericity
axis~\cite{Bjorken:1969wi} of the \Bz{} candidate and that of the
remaining particles in the event.  While $|\cos\theta^*_S|$ is highly
peaked near 1 for $\epem\to\qqbar$ events, it is nearly uniformly
distributed for \BB{}.  We require $|\cos\theta^*_S|<0.8$, eliminating
$83\%$ of the background.  In addition, we include in the fit a Fisher
discriminant variable, which is defined as ${\cal F} = 0.53 - 0.60 L_0
+ 1.27 L_2$, where $L_j\equiv\sum_i |{\bf p}^*_i| |\cos
\theta^*_i|^j$, ${\bf p}^*_i$ is the momentum of particle $i$ and
$\theta^*_i$ is the angle between ${\bf p}^*_i$ and the sphericity
axis of the \Bz{} candidate.

We use a neural network (NN) to determine the flavor of the $\Btag$
meson from kinematic and particle identification
information~\cite{ref:sin2betaPRL02}. Each event is assigned to one of
five mutually exclusive tagging categories, designed to combine flavor
tags with similar performance and \deltat\ resolution.  We
parameterize the performance of this algorithm in a data sample
($B_{\rm flav}$) of fully reconstructed $\Bz\to D^{(*)-}
\pip/\rho^+/a_1^+$ decays. The average effective tagging efficiency
obtained from this sample is $Q = \sum_c \epsilon_S^c
(1-2w^c)^2=0.288\pm 0.005$, where $\epsilon_S^c$ and $w^c$ are the
efficiencies and mistag probabilities, respectively, for events tagged
in category $c$. For the background the fraction of events
($\epsilon_B^c$) and the asymmetry in the rate of $\Bz$ versus $\Bzb$
tags in each tagging category are extracted from the fit to the data.

We compute the proper time difference \deltat{} from the known boost
of the \epem{} system and the measured $\deltaz=\zrec-\ztag$, the
difference of the reconstructed decay vertex positions of the
\Bztokspiz{} and \Btag{} candidate along the boost direction ($z$).  A
description of the inclusive reconstruction of the \Btag{} vertex is
given in \cite{ref:Sin2betaPRD}.  For the \Bztokspiz{} decay, where no
charged particles are present at the decay vertex, we exploit the fact
that the flight distance of the $\B$ meson transverse to the beam
direction (\unit[$\sim30$]{\microns}) is small compared to the flight
length along the beam (\unit[$\sim260$]{\microns}). We then determine
the decay point from the intersection of the \KS{} trajectory with the
interaction region by constraining the \B{} vertex to the interaction
point (IP) in the transverse plane.  The position and size of the
interaction region are determined on a run-by-run basis from the
spatial distribution of vertices from two-track events.  The
uncertainty in the IP position, which follows from the size of the
interaction region (about \unit[200]{$\mu$m} horizontal and
\unit[4]{$\mu$m} vertical), is combined with the RMS of the transverse
\B{} flight length distribution to assign an uncertainty to the IP
constraint.

Simulation studies indicate that the vertexing procedure provides an
unbiased estimate of \zrec{}. The per-event estimate of the $\deltat$
error reflects the strong dependence of the \zrec{} resolution on the
$\KS$ flight direction and the number of SVT layers traversed by its
decay daughters. For the $37\%$ of events where both tracks include at
least one hit in the inner three SVT layers (at radii from
\unit[$3.2$]{cm} to \unit[$5.4$]{cm}), the mean \deltat\ resolution is
comparable to that of decays for which the vertex is directly
reconstructed from charged particles originating at the $B$ decay
point~\cite{ref:Sin2betaPRD}.  If both tracks have hits in the outer
two SVT layers (at radii \unit[$9.1$]{cm} to \unit[$14.4$]{cm}) but
one of the tracks has no hits in the inner three layers ($\sim27\%$ of
the events), the resolution is nearly two times worse. The remaining
events provide poor $\deltat$ measurements.  For these events and for
events with \unit[$\sigma_{\deltat}>2.5$]{ps} or
\unit[$|\deltat|>20$]{ps}, we do not include $\deltat$ information in
the fit.  However, we account for the contribution of these events in
the measurement of \ckspiz.

We obtain the PDF for the time-dependence of signal decays from the
convolution of Eq.~\ref{eqn:td} with a resolution function ${\cal
  R}(\delta t \equiv \deltat -\deltat_{\rm true},\sigma_{\deltat})$.
The resolution function is parameterized as the sum of a `core' and a
`tail' Gaussian, each with a width and mean proportional to the
reconstructed $\sigma_{\deltat}$, and a third Gaussian centered at
zero with a fixed width of \unit[$8$]{ps}~\cite{ref:Sin2betaPRD}.  We
have verified in simulation that the parameters of ${\cal R}(\delta t,
\sigma_{\deltat})$ for \Bztokspiz\ decays are similar to those
obtained from the $B_{\rm flav}$ sample, even though the distributions
of $\sigma_{\deltat}$ differ considerably. Therefore, we extract these
parameters from a fit to the $B_{\rm flav}$ sample.  We find that the
\deltat{} distribution of background candidates is well described by a
delta function convolved with a resolution function with the same
functional form as used for signal events. The parameters of the
background function are determined in the fit.

To extract the CPV asymmetries we maximize the logarithm of the
likelihood function
{\small\begin{eqnarray*}
  \lefteqn{{\cal L}(\sf,\cf,N_S,N_B,f_S,f_B,\vec{\alpha}) =\frac{e^{-(N_S+N_B)}}{(N_S+N_B)\,!}  \times}  & 
  \mbox{}\rule{50em}{0pt} \nonumber\\
  \lefteqn{\prod_{i \in \mathrm{w/\,} \deltat}
  \left[ N_S f_S \epsilon^{c}_S{\cal P}_S(\vec{x}_i,\vec{y}_i;\sf,\cf) + 
    N_B f_B \epsilon^{c}_B {\cal P}_B(\vec{x}_i,\vec{y}_i;\vec{\alpha}) \right] \times}  & \mbox{}  \nonumber\\ 
  \lefteqn{\prod_{i \in  \mathrm{w/o\,} \deltat}
  \left[ N_S (1-f_S) \epsilon^{c}_S {\cal P}'_S(\vec{x}_i;\cf) + 
    N_B (1-f_B) \epsilon_B^{c} {\cal P}'_B(\vec{x}_i;\vec{\alpha}) \right],} & \mbox{} \nonumber
\end{eqnarray*}}
\noindent{}where the second (third) factor on the right-hand side is the
contribution from events with (without) $\deltat$
information. The probabilities ${\cal P}_S$ and ${\cal P}_B$ are
products of PDFs for signal ($S$) and background ($B$) hypotheses
evaluated for the measurements
$\vec{x}_i=\{\mes,\DeltaE,\fish,\text{tag},\text{tagging category}\}$
and $\vec{y}_i=\{\deltat,\sigma_{\deltat}\}$. Along with the CPV
asymmetries \sf\ and \cf, the fit extracts the yields $N_S$ and $N_B$,
the fractions of events with $\deltat$ information $f_S$ and $f_B$,
and the parameters $\vec{\alpha}$ which describe the background PDFs.

Fitting the data sample of 4179 $\Bztokspiz$ candidates, we find
\mbox{$N_S=122\pm 16$} signal decays with \mbox{$\skspiz =
  0.48^{+0.38}_{-0.47} \pm 0.06$} and \mbox{$\ckspiz =
  0.40^{+0.27}_{-0.28} \pm 0.09$}, where the uncertainties are
statistical and systematic, respectively. The estimated number of
signal decays is consistent with our measurement of the branching
fraction~\cite{Aubert:2003sg}.  The result for $\ckspiz$ is consistent
with a fit that does not employ $\deltat$ information.  Fixing
\mbox{$\ckspiz=0$} we obtain \mbox{$\skspiz=0.41^{+0.41}_{-0.48}\pm
  0.06$}. The evaluation of the systematic uncertainties is described
below.

\begin{figure}[!tbp]
\begin{center}
\includegraphics[width=0.9\linewidth]{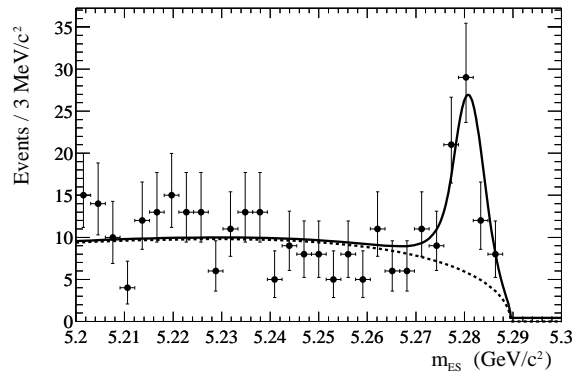}
\caption{Distribution of $\mes$ for events enhanced
  in signal decays.  The dashed and solid curves represent the
  background and signal-plus-background contributions, respectively,
  as obtained from the maximum likelihood fit.}
\label{fig:prplots}
\end{center}
\end{figure}

\begin{figure}[!tbp]
\begin{center}
\includegraphics[width=0.9\linewidth]{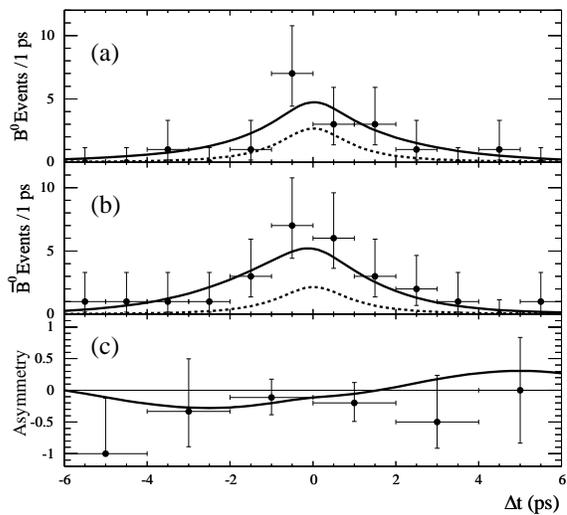}
\end{center}
\caption{
  Distributions of $\deltat$ for events enhanced in signal decays with
  $B_{\rm tag}$ tagged as (a) $\Bz$ or (b) $\Bzb$, and (c) the
  asymmetry ${\cal A}_{\KS\piz}(\deltat)$.
  The dashed and solid curves represent the fitted background and
  signal-plus-background contributions, respectively, as obtained from
  the maximum likelihood fit. The asymmetry projection corresponds to
  approximately $36$ signal and $25$ background events.}
\label{fig:dtplot}
\end{figure}

Figure~\ref{fig:prplots} shows the \mes{} distributions for a
signal-enhanced sample. The event selection is based on a likelihood
ratio $R={\cal P}_S/({\cal P}_B+{\cal P}_S)$ calculated without the
displayed observable. The dashed and solid curves indicate background
and signal-plus-background contributions, respectively, as obtained
from the fit, but corrected for the selection on $R$.
Figure~\ref{fig:dtplot} shows distributions of $\deltat$ for $\Bz$-
and $\Bzb$-tagged events, and the asymmetry ${\cal
  A}_{\KS\piz}(\deltat) = \left[N_{\Bz} -
  N_{\Bzb}\right]/\left[N_{\Bz} + N_{\Bzb}\right]$ as a function of
$\deltat$, also for a signal-enhanced sample.

In order to investigate possible biases introduced in the CPV
measurements by the IP-constrained vertexing technique, we examine
\Bztojpsiks{} decays in data, where $\jpsi\to\mup\mun$ and $\jpsi\to
\epem$.  In these events we determine $\deltat$ in two ways: by fully
reconstructing the $\Bz$ decay vertex using the trajectories of
charged daughters of the $\jpsi$ and the $\KS$ mesons, or by
neglecting the $\jpsi$ contribution to the decay vertex and using the
IP constraint and the \KS{} trajectory only. This study shows that
within statistical uncertainties the IP-constrained $\deltat$
measurement is unbiased with respect to the more established technique
and that the obtained values of $S_{\jpsi\KS}$ and $C_{\jpsi\KS}$ are
consistent. A similar study of $\Bpm\to\KS\pipm$ events, where the
$\pipm$ contribution to the decay vertex has been replaced by the IP
constraint, yields $S_{\KS\pipm}= 0.13 \pm 0.19 $ and $C_{\KS\pipm}=
0.06 \pm 0.11$, which is consistent with the expectation
$S_{\KS\pipm}=0$ and our previous measurement of the charge
asymmetry~\cite{Aubert:2003sg}.  We also find that the $\Bz$ lifetime
measured in \Bztokspiz{} decays and in IP-constrained \Bztojpsiks{}
decays agrees with the world average~\cite{Hagiwara:fs}.

To quantify possible systematic effects we examine large samples of
simulated \Bztokspiz\ and \Bztojpsiks\ decays. We employ the
difference in resolution function parameters extracted from these
samples to evaluate uncertainties due to the use of the resolution
function ${\cal R}$ extracted from the $B_{\rm flav}$ sample. We
assign a systematic uncertainty of $0.03$ on \skspiz{} and $0.02$ on
\ckspiz{} due to the uncertainty in ${\cal R}$. We compare fits to a
large sample of simulated nominal and IP-constrained \Bztojpsiks\ 
events to account for any potential bias due to the vertexing
technique. This latter study yields the difference $\delta
S_{\jpsi\KS}=0.04$, which we assign as the dominant systematic
uncertainty on \skspiz.  We include a systematic uncertainty of $0.03$
on \skspiz{} and $0.01$ on \ckspiz{} to account for a possible
misalignment of the SVT. We consider large variations of the IP
position and resolution, which we find to have negligible impact. We
assign a systematic uncertainty of $0.09$ to \ckspiz{} due to possible
asymmetries in the rate of \Bz{} versus \Bzb{} tags in background
events. Finally, we include a systematic uncertainty of $0.02$ on both
\skspiz{} and \ckspiz{} to account for imperfect knowledge of the PDFs
used in the fit.

In summary, we have performed a measurement of the time-dependent CPV
asymmetries in \Bztokspiz. These results supersede our previous
measurement of \ckspiz{}~\cite{Aubert:2003sg}, which only relied on
time-integrated observables, and introduce the first measurement of
\skspiz.

\par
We are grateful for the excellent luminosity and machine conditions
provided by our \pep2\ colleagues, 
and for the substantial dedicated effort from
the computing organizations that support \babar.
The collaborating institutions wish to thank 
SLAC for its support and kind hospitality. 
This work is supported by
DOE
and NSF (USA),
NSERC (Canada),
IHEP (China),
CEA and
CNRS-IN2P3
(France),
BMBF and DFG
(Germany),
INFN (Italy),
FOM (The Netherlands),
NFR (Norway),
MIST (Russia), and
PPARC (United Kingdom). 
Individuals have received support from the 
A.~P.~Sloan Foundation, 
Research Corporation,
and Alexander von Humboldt Foundation.


\begin{thebibliography}{99}

\bibitem{BaBarSin2betaObs}
 B. Aubert {\em et al.} [\babar{} Collaboration], \jprl{\bf 87}, 091801 (2001).

\bibitem{BelleSin2betaObs}
  K. Abe {\em et al.} [BELLE Collaboration] , \jprl{\bf 87}, 091802 (2001).

\bibitem{CKM}
  N.~Cabibbo, \jprl {\bf 10}, 531 (1963);
  M.~Kobayashi and T.~Maskawa, \progtp{\bf 49}, 652 (1973).

\bibitem{Grossman:1996ke}
Y.~Grossman and M.~P.~Worah,
\plb{\bf 395}, 241 (1997).

\bibitem{Ciuchini:1997zp}
M.~Ciuchini, E.~Franco, G.~Martinelli, A.~Masiero and L.~Silvestrini,
\jprl{\bf 79}, 978 (1997).

\bibitem{ref:cc} Unless explicitly stated, conjugate decay modes are
assumed throughout this paper.

\bibitem{Abe:2003yt}
K.~Abe {\it et al.}  [Belle Collaboration],
\jprl{\bf 91}, 261602 (2003).

\bibitem{Aubert:2003bq}
B.~Aubert {\it et al.}  [\babar{} Collaboration],
\jprl{\bf 91}, 161801 (2003).

\bibitem{ref:browderlp03} T. E. Browder, to appear in the Proceedings
  of the 2003 Lepton-Photon Conference,
  Fermilab, hep-ex/0312024.

\bibitem{ref:HFAG}
Average computed by the Heavy Flavor Averaging Group, J.~Alexander,
{\it et. al.}.
A. Bornheim \etal{} [CLEO Collaboration], hep-ex/0302026 (2003). 
T. Tomura (for the Belle Collaboration), hep-ex/0305036 (results presented at Moriond Hadronic 2003).
B.~Aubert \etal{} [\babar{} Collaboration], hep-ex/0312055.

\bibitem{Gronau:2003kx}
M.~Gronau, Y.~Grossman and J.~L.~Rosner,
\plb{\bf 579}, 331 (2004).

\bibitem{ref:babar}
B. Aubert {\em et al.} [\babar{} Collaboration] , \nima{479}, 1 (2002).

\bibitem{ref:Sin2betaPRD}
B. Aubert {\em et al.} [\babar{} Collaboration] ,
\jprd{\bf 66}, 032003 (2002).

\bibitem{Hagiwara:fs}
K.~Hagiwara {\it et al.} [Particle Data Group Collaboration],
\jprd{\bf 66}, 010001 (2002).

\bibitem{Bjorken:1969wi}
J.~D.~Bjorken and S.~J.~Brodsky,
\jprd{\bf 1}, 1416 (1970).

\bibitem{ref:sin2betaPRL02}
B.~Aubert {\it et al.} [\babar{} Collaboration] ,
\jprl{\bf 89}, 201802 (2002).

\bibitem{Aubert:2003sg}
B.~Aubert {\it et al.}  [\babar{} Collaboration],
hep-ex/0312055, submitted to \jprlBase{}.

\end{thebibliography}
\end{document}